\input{aipcheck}

\documentclass[
    ,final            
  ]
  {aipproc}

\layoutstyle{8x11single}

\newcommand{\ea}{{et al.}}

\def\asa{A\&A}

\begin{document}

\title{The XMM-{\fontshape{it}\selectfont{}Newton} view of  GRS 1915+105 during a ``plateau''}

\classification{97.60.-s, 97.80.Jp, 98.70.Qy, 98.62.Mw, 98.62.Nx
}
\keywords      {Black hole physics -- Accretion, accretion discs --   
X-rays: binaries -- X-rays: individuals: GRS\,1915+105 }

\author{A.\ Martocchia\thanks{{\it Present address:} UPS / Centre d'Etude Spatiale des Rayonnements, 9 Av. du Colonel Roche, F--31028 Toulouse, France}}{
  address={CNRS / Observatoire Astronomique de Strasbourg, 11 Rue de l'Universit\'e,    
F--67000 Strasbourg, France}
}

\author{G.\ Matt}{
address={Dipartimento di Fisica, Universit\`a degli Studi ``Roma Tre'',   
Via della Vasca Navale 84, I--00146 Roma, Italy}
}

\author{T.\ Belloni}{
address={INAF / Osservatorio Astronomico di Brera, via E. Bianchi 46, I--23807 Merate, Italy}
}

\author{M.\ Feroci}{
address={INAF / IASF, Area di Ricerca di Tor Vergata, Via Fosso del Cavaliere 100,    
I--00133 Roma, Italy }
}

\author{V.\ Karas}{
address={Astronomical Institute, Academy of Sciences,    
Bo\v{c}n\'{\i}~II, CZ--140\,31~Prague, Czech Republic }
}

\author{G.\ Ponti}{
address={Institute of Astronomy, Madingley Road, Cambridge CB3 0HA, United Kingdom }
,altaddress={Dipartimento di Astronomia, Universit\`a di Bologna, via Ranzani 1, I--40127 Bologna, Italy }
,altaddress={INAF / IASF Sezione di Bologna, via Gobetti 101, I--40129 Bologna, Italy }
}

\begin{abstract}
Two XMM-{\fontshape{it}\selectfont{}Newton} observations of the black-hole binary GRS1915+105 were triggered in 2004 (April 17 and 21), during a long "plateau" state of the source. We analyzed the data collected with EPIC-pn in Timing and Burst modes, respectively. No thermal disc emission is required by the data; the spectrum is well fitted by four components: a primary component (either a simple power law or thermal Comptonization) absorbed by cold matter with abundances different than those of standard ISM; reprocessing from an ionized disc; emission and absorption lines; and a soft X-ray excess around 1 keV. The latter is not confirmed by RGS (which were used in the second observation only); if real, the excess could be due to reflection from the optically thin, photoionized plasma of a disc wind, in which case it may provide a way to disentangle intrinsic from interstellar absorption. Indeed, the former is best traced by the higher abundances of heavier elements, while an independent estimate of the latter may be given by the value we get for the disc wind component only, which roughly coincides with what is found for lower-Z species.
\end{abstract}

\maketitle


\section{Introduction}

GRS\,1915+105 is a well-known black-hole (BH) binary showing superluminal jets and
with very peculiar variability properties (for a recent review on this source see Fender \& Belloni, 2004). Due to very large obscuration, the spectral type of GRS\,1915+105's companion (a K-M III star) was discovered lately, via infrared observations, which also helped to finally determine the mass of the central compact object, which has been constrained to $M_{\rm c}=14\pm4 M_\odot$ (Greiner \ea, 2001). 

A XMM-{\it Newton} ToO observation of GRS\,1915+105 was proposed in AO2.  The observation was intended to be triggered by the occurrence of a ``plateau'' state of the source similar to that observed during the {\it Beppo}SAX 1998 observation, when relativistic Fe lines were observed (Martocchia \ea\ 2002, 2004); this was necessary also in order to have the source in a less dramatic variability state, and at a lower flux level to minimize technical problems due to instrumental pile-up and telemetry. The observation was triggered in April 2004, divided into two parts: OBS1 (April 17) and OBS2 (April 21; see Table~\ref{tab} for details).   

\section{Results}

\begin{table}
    \caption{The main parameters of the two observations and EPIC-pn data best fits. 
    The elemental abundancies are in units of 10$^{22}$ cm$^{-2}$. The symbol ($^*$)
    indicates a frozen parameter. For 
    more details, including a full description of the spectral fitting procedure, see 
    Martocchia \ea\ (2006).  }\vspace{1em}
    \renewcommand{\arraystretch}{1.2}
    \begin{tabular}[h]{|c|c|c|}
    \hline
				& {\bf OBS1} 		& {\bf OBS2} \\ 
\hline
{\bf Date}			& 2005-04-17 		& 2005-04-21 \cr   
{\bf Obs. mode}		& Timing 			& Burst \cr   
{\bf duration} [s]		& 20681 			& 25652 \cr   
{\bf RGS mode}		& OFF 			& HCR \cr   
      \hline
{\bf Power law $\Gamma$} & 1.686$^{+0.008}_{-0.012}$ & 2.04$^{+0.01}_{-0.02}$ \cr   
{\bf Cold {\fontshape{it}\selectfont{}intrinsic} absorption} & ~ & \cr   
$N_{\rm H,He,C,N,O}$ 
				& 1.60$^{+0.17}_{-0.29}$ & 1.98$^{+0.02}_{-0.02}$ \cr   
$N_{\rm Ne,Na}$ 
				& 7.46$^{+0.31}_{-0.72}$ & same ratio as OBS1$^*$ \cr   
$N_{\rm Mg,Al}$ 
				& 7.57$^{+0.54}_{-0.16}$ & same ratio as OBS1$^*$ \cr   
$N_{\rm Si}$ 
				& 5.70$^{+0.07}_{-0.12}$ & same ratio as OBS1$^*$ \cr   
$N_{\rm S}$ 
				& 4.69$^{+0.07}_{-0.69}$ & same ratio as OBS1$^*$ \cr   
$N_{\rm Cl,Ar,Ca,Cr}$ 
				& 11.0$^{+1.3}_{-1.6}$      & same ratio as OBS1$^*$ \cr   
$N_{\rm Fe,Co,Ni}$ 
				& 11.7$^{+0.2}_{-0.2}$      & same ratio as OBS1$^*$ \cr   
\hline 
{\bf Emis. line (Si{\sc xiii} ?)} 	& ~ & \cr   
$E_l$ [keV]		& 1.846$^{+0.006}_{-0.005}$ 	 	& -- \cr   
$F_l$ [10$^{-3}$ ph cm$^{-2}$ s$^{-1}$]
				& 13.5$^{+1.5}_{-0.7}$  			& -- \cr   
$EW$ [eV] 		&  19  						& $<$5 \cr    
{\bf Emis. line (Si{\sc xii+xiv} ?)} 	& ~ & \cr 
$E_l$ [keV]		& 2.244$^{+0.007}_{-0.010}$ 	 	& -- \cr   
$F_l$ [10$^{-3}$ ph cm$^{-2}$ s$^{-1}$]
				& 3.62$^{+0.46}_{-0.62}$			& -- \cr   
$EW$ [eV] 		& 8  							& $<$2  \cr   
\hline 
{\bf Iron lines} 		& ~ & \cr   
{\bf He-like}: $E_c=6.7$ keV$^*$ 		& ~ & \cr 
$r_i/r_g$ 			& 580$^{+210}_{-120}$ 	& -- \cr   
$F_l$ [10$^{-3}$ ph cm$^{-2}$ s$^{-1}$]			
				& 2.37$^{+0.18}_{-0.18}$ 	& -- \cr   
$EW$ [eV]		& 28  				& $<$10  \cr   
{\bf H-like}: $E_c=6.96$ keV$^*$ 	& ~ & \cr    
$r_i/r_g$ 			& 320$^{+80}_{-60}$ 	& -- \cr   
$F_l$ [10$^{-3}$ ph cm$^{-2}$ s$^{-1}$]
				& 2.20$^{+0.19}_{-0.21}$	& -- \cr   
$EW$ [eV]		& 28  				& $<$10  \cr   
Absorption line: 			& ~ & \cr  
$E_l$ [keV]		& 6.95$^{+0.01}_{-0.03}$ & 6.98$\pm0.02$  \cr   
$F_l$ [10$^{-3}$ ph cm$^{-2}$ s$^{-1}$]
				& -0.79$^{+0.02}_{-0.01}$ 	& \cr   
$EW$ [eV] 		& -9  						& -50  \cr    
$\sigma$ [eV]		& 1$^*$ 					& 110$\pm30$  \cr   
\hline 
{\bf Reflection} 		& ~ & \cr   
$R/2\pi$ 			& 0.35$^{+0.02}_{-0.02}$ & 1.69$^{+0.16}_{-0.04}$ \cr   
$A_{\rm Fe}$ 		& 5.2$^{+0.7}_{-1.9}$ 	& 5.2$^*$ \cr   
$\xi$ [erg cm s$^{-1}$] 
				& 940$^{+190}_{-80}$ 	& 3300$^{+600}_{-600}$ \cr   
$r_i/r_g$ 			& 320$^{+80}_{-60}$ 	& $<$20 \cr   
\hline
$F_{\rm 2-10~keV}$ [$10^{-8}$ erg cm$^{-2}$ s$^{-1}$] 
				& $\sim 0.6$ (unabs: $\sim 0.87$) & $\sim 0.66$ (unabs: $\sim 1.07$) \cr  
$\chi^2$/d.o.f. 		& 317.5/227 			& 248.2/219 \cr   
      \hline 
      \end{tabular}
    \label{tab}
\end{table}

\begin{figure}   
\hspace{-0.5cm}   
\includegraphics[width=0.55\textwidth]{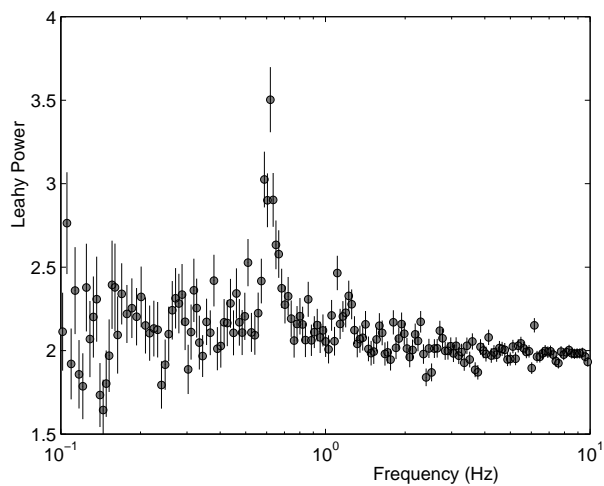}
\caption{Power spectrum in the 0.4 -- 13.0 keV band of the OBS2 EPIC-pn data. }   
\label{figqpo}   
\end{figure}

We succedeed at both a) observing the source in a well-defined,  stable physical/spectral state and b) collecting EPIC-pn useful data, only marginally corrupted by telemetry problems.    
In both observations the source has been caught in a ``plateau'' state, which we identify with 
the conventional ``C'' spectral state / $\chi$ variability mode as defined by Belloni \ea\ (2000; see also Fender \& Belloni, 2004). It shows a QPO at $\sim 0.6$ Hz -- i.e. what is expected    
in ``plateau'' intervals when the frequency vs. spectral hardness correlation is taken into account -- with a possible harmonic signal at 1.2 Hz (Fig.~\ref{figqpo}).   

We adopted a power law continuum model, which mimics emission by a hot corona or Comptonized thermal emission e.g. from the jet basis; however, an optically thick reflector is required to account for the smeared edge at $\sim 7$ keV, with a covering ratio of $\sim 0.4 \div 1.7 \times 2\pi$
 in the two observations respectively. 
The latter component yields evidence of an accretion    
disc being present, or just optically thick, only at quite large distance from the central compact object, al least in the first observation ($r_i/r_g > 300$ in OBS1, $\sim 20$ in OBS2).  
The relatively large amount of the reflection components implies that the primary 
X--ray emitting region should have a size at least comparable to the inner disc radius.  
That the disc is truncated, i.e. not present in the innermost part, is suggested also by the non-detection of thermal disc emission.    
    
Several line residuals are superimposed on the modeled continuum. Part of these may be due to calibration uncertainties, especially at the energies where changes in the EPIC effective area take place (e.g. 1--3 keV). However, we found clear evidence of ionized iron emission around $\sim 7$ keV: data are well fitted with two ionized Fe K$\alpha$ lines, possibly affected by mild relativistic broadening (being produced far away from the BH event horizon), plus a narrow absorption feature at $\sim 6.95$ keV.   
  
Finally, we register the puzzling presence of an intense, broad excess around 
1 keV in EPIC-pn data; the RGS spectrum does not confirm this, showing instead 
a fast decline, and no apparent features. Several hypotheses, which can be invoked to explain the RGS--pn discrepancy, are discussed in Martocchia \ea\ (2006). 
Assuming as a working hypothesis that the 1 keV excess is real 
(which would imply RGS calibration problems, an admittedly unsubstantied  
assumption at the moment), 
it could be satisfactorily explained in terms of reflection by an optically thin wind.   
The excess is indeed well fitted with a power law plus a line, unobscured by material intrinsic to the system. The centroid energy of the gaussian line ($\sim$0.97 keV), its width (90 eV), and its EW against the reflected continuum (5.6 keV), point to a blend of Ne K and Fe L lines. The value of the equivalent H column density (as given by the OBS1 best fit, and frozen while fitting OBS2 data) results to be interestingly similar to the value of the obscuration by low Z elements (H, He, C, N, O) at the source core -- $N_{\rm H} \sim 1.6\times10^{22}$ cm$^{-2}$: if the disc wind hypothesis is true, 
this may therefore be taken as an upper limit to the {\it interstellar} matter column density. 
This value matches well with the expected galactic absorption in that direction (Dickey \& Lockman 1990).   

\begin{figure}   
\includegraphics[width=7.0cm,angle=-90]{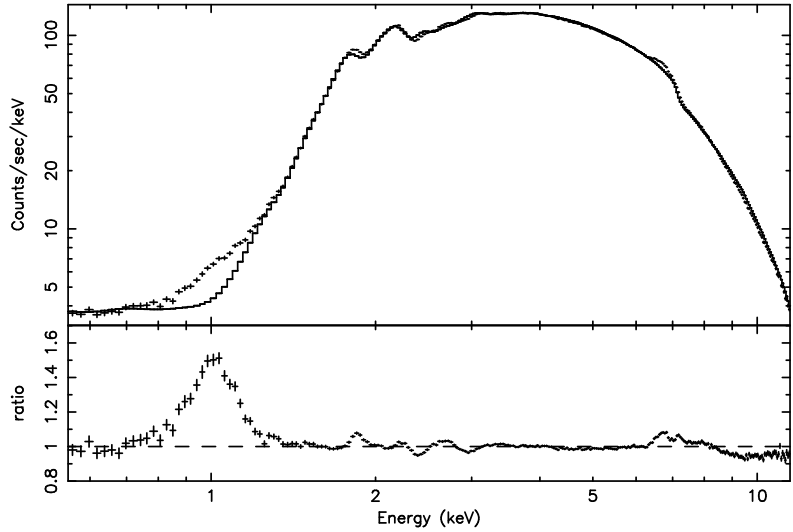}
\caption{The OBS1 spectrum and data/model ratio, after fitting to a simple power law plus cold absorption model, clearly show the most significant residuals (see text and Table 1).}   
\label{gio1pl}   
\end{figure}   
\begin{figure}   
\vspace{-1cm}  
\includegraphics[height=8.8cm,width=7.0cm,angle=-90]{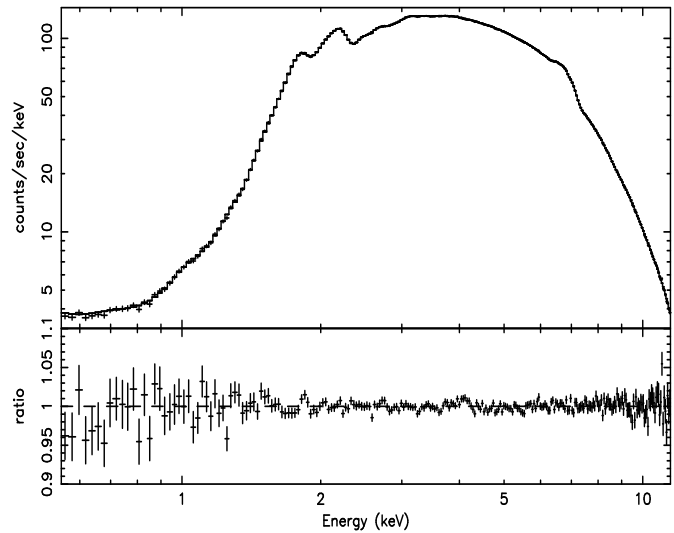}
\caption{The OBS1 spectrum and data/model ratio for the best fit model (see text and Table).}   
\label{bestfit}   
\end{figure}   

On the other hand, a significant fraction of the absorber must  be {\it local} to the source.   
We adopted a variable absorption model ({\sc varabs} in {\sc XSPEC}), assuming neutral matter    
and grouping the elements on the base of both physical and practical considerations: elements which have probably a common origin, but also elements which are not very abundant (and therefore cannot be easily  measured independently one from the other) with very abundant ones (e.g. Co and Ni with Fe: see the results in Table~\ref{tab}).   
A significant overabundance of the heavier elements with respect to the lighter ones is apparent, which suggests that a significant fraction of the absorber, traced by heavier species, is local to the source. Clearly, the intrinsic absorption may be subject to substantial changes on longer timescales, as already observed with {\it Rossi}-XTE in correspondence of similar ``plateaux'' (Belloni \ea\ 2000).  \\ 

Line features are less apparent in OBS2 than in OBS1; a 6.4 keV iron emission line is 
marginally found, with an EW  of 6$\pm$4 eV. The steeper spectrum can be due to the more efficient cooling due to the increase of soft disc photons. Regarding the 1 keV excess, which still persists, it can be again well fitted by a power law plus an emission line, the latter with a flux slightly larger than that obtained in the Timing-mode observation. 

The results of OBS2 (see Table~\ref{tab}) are consistent with a picture in which the disc is more extended toward the compact object, and more ionized. However, the estimates of the disc radii must be taken with caution, since they are now determined via the reflection component only. 

Moreover, the OBS2 spectrum at the higher energies can be at least partly affected by Burst-mode calibration problems. A deficit of photons above 6$\div$7 keV, i.e. similar to what we find, is indeed present also in the Burst mode observation of the Crab Nebula  (Kirsch et al. 2005), a source where of course no significant reflection component is expected.

Sala et al. (2006) analized XMM-{\it Newton} data of GRS\,1915+105 taken in Burst mode again just a few days after our observations. They detected the 1 keV excess anew. These authors are mainly concerned with the calibration problems related to Charge Transfer Efficiency, which however do not really affect GRS\,1915+105 data and thus cannot help to cure the 1 keV excess.    \\
    
A priori, some of the features in both spectra may be affected by dust halo scattering, too. 
We cannot  check this hypothesis with our data, given the lack of imaging capabilities of timing  
modes (but see Greiner et al., 1998, on the issue); 
however, while spectral modelling of such effects is not easy, they would not help 
explaining the 1 keV excess entirely: the halo spectrum, normalized to the source 
spectrum in Cyg\,X-2 by Costantini et al. (2005), is much broader than our 1 keV excess. 

Finally, in order to try disentagle the different spectral components we used the $rms$ vs. $E$ method by Ponti \ea\ (2004). The resulting $rms$ is lower than 0.1 all over the energy band, i.e. all spectral components are compatible with being constant, on timescales bigger than $\sim 100$ s, during the observation.


\begin{theacknowledgments}
AM wishes to thank the French Space Agency CNES and 
the CNRS Group of Research ``Ph\'enom\`enes Cosmiques  
de Haute Energie'' for financial support, as well as the Astronomical  
Institute of the Czech Academy of Sciences for the very nice hospitality. 
TB acknowledges partial support by grants INAF-PRIN 2002 and  
MIUR-PRIN 2003027534\_004. 
VK acknowledges the grant ref. GAAV IAA\,300030510. 
GP thanks the European commission under the Marie Curie Early Stage Research  
Training programme for support. 
\end{theacknowledgments}



\bibliographystyle{aipprocl} 



\end{document}